\documentclass[a4paper,11pt]{article}
\usepackage{jcappub} 
\usepackage{lineno}

\usepackage{subcaption}
\usepackage{color}
\newcommand{\red}{\color{black}}
\newcommand{\black}{\color{black}}

\title{\boldmath TeV to PeV neutrinos from AGN coronae}

\author{Simon Sotirov}
\affiliation{Physics Department, M.V. Lomonosov Moscow State University, 1-2 Leninskie Gory,  Moscow 119991, Russia}

\affiliation{Institute for Nuclear
Research of the Russian Academy of Sciences, 60th October Anniversary Prospect 7a, Moscow 117312, Russia}

\emailAdd{sotirov.sa19@physics.msu.ru}

\abstract{In this paper, we attempt to explain the TeV - sub PeV neutrinos observed by IceCube assuming that their sources are active galactic nuclei (AGN). The results are obtained in the model where the thermal accretion disc emits in the UV-optical range inside the hot electron plasma cloud. We focus on the analytical solution for the comptonization problem obtained from consideration at the microscopic level and try to avoid fitting the coronal spectrum from spectral observations. Using the Monte-Carlo approach to model photopion interactions in the central regions of AGN and then after taking into account the cosmological evolution it is shown that within the framework of this approach it is possible to describe both $\sim$ 100 TeV and sub PeV neutrinos from AGN taking into account only photohadronic interactions.}

\begin{document}
\maketitle
\flushbottom

\section{Introduction}
\label{sec:intro}

High-energy astrophysical neutrinos are one of the important attributes of modern astrophysics because they can contain information about the structure and processes occurring in astrophysical sources. IceCube observes the diffuse astrophysical neutrino flux of $\sim 10^{-8}$ GeV cm$^{-2}$ s$^{-1}$ sr$^{-1}$ \cite{Naab:2023xcz} in the energy range from $\sim$ 1 TeV to $\sim$ 10 PeV with high statistical significance. This provides impetus for the construction of theoretical models and possible explanation. However, due to the large error in reconstructing the direction of arrival of neutrino events, it is not always possible to determine the astrophysical source unambiguously \cite{Palladino:2020jol}.

Many extragalactic candidates have been proposed in the literature as neutrino sources: AGN \cite{Kalashev:2014vya, Inoue:2019fil, Murase:2019vdl, Padovani:2024ibi,Abbasi:2024ofy}, starburst galaxies \cite{Loeb:2006tw}, gamma ray bursts \cite{Waxman:1997ti}. Regardless of the source, the production of astrophysical neutrinos requires the presence of high-energy protons or nuclei. General constraints can be imposed on the maximal energy of accelerated protons  \cite{Hillas:1984ijl, Ptitsyna:2008zs, Sotirov:2022uqk} - one of the parameters on which the neutrino spectrum depends. Due to the high luminosity of the AGN, stronger constraints may be imposed on the maximum proton energy \cite{Sotirov:2022uqk}.

 In this paper, we are interested in AGN as the main sources. The calculation of the neutrino spectrum, which includes modeling of photohadronic interactions in the central regions of the AGN, was performed in \cite{Kalashev:2014vya}. In the work the model is considered where the only source of photons inside the AGN is the Shakura-Sunyaev accretion disk \cite{Shakura:1972te}, which is bright in the ultraviolet - optical range (the so-called big blue bump). Because the accretion disk photons have low energies (of about $\sim$ 10 eV), the predicted flux has the shape of a narrow bump while the IceCube spectrum is broader. 

However, AGN are also bright in the X-ray range \cite{Fabian:1992rt, Ueda:2014tma}  which is due to the presence of the so-called corona, which must be taken into account in the calculations. Based on observational data, empirical relations and consideration of specific acceleration mechanisms, models were proposed to calculate the spectrum of particles from AGN in \cite{Inoue:2019fil, Murase:2019vdl}, where the corona is of great importance.

The detection of $\sim$ 10 TeV neutrinos from AGN by IceCube \cite{IceCube:2018cha, Plavin:2020mkf, IceCube:2022der} demonstrates that thermal radiation from the accretion disk cannot efficiently produce neutrinos at these energies see e.g. \cite{Murase:2019vdl, Neronov:2025cfc}. However, the present study shows that accounting for coronal emission resolves the issue, enabling effective neutrino production across a broad energy range.
To do this, we modify the spectrum of the target source photons and model the propagation of protons along an accretion disk axis using the Monte Carlo approach. Then, using the numerical code based on the solution of the transport equations, we find the final diffuse neutrino spectrum from AGN, taking into account their cosmological evolution. Finally, we normalize the calculated neutrino spectrum to the IceCube diffuse flux.

While the observed astrophysical neutrino flux has multiple components—including a well-established Galactic contribution \cite{Kovalev:2022izi, IceCube:2023ame, Baikal-GVD:2024kfx}—current statistics remain insufficient to disentangle all sources definitively. Only a fraction of the total flux may be linked to AGN \cite{McDonough:2023ngk}. For this reason, the present work does not aim to fit the observed neutrino spectrum. 

The paper is organized as follows. Section \ref{2}  describes in detail the disk-corona model, namely the spectrum of the corona, the disk, and the neutrino production mechanism. Section \ref{3} presents the procedure for modeling proton propagation in the central regions of AGN. Section \ref{4} presents the final results and discussion.

\section{Disc-corona model}
\label{2}
In a simplified model, the AGN is a supermassive black hole (SMBH) surrounded by an accretion disk, which is approximately described by a thermal spectrum with a temperature of about $\sim$ 10 eV \cite{Shakura:1972te}. From the core of the AGN the relativistic jet is erupted perpendicular to the accretion disk. Charged particles in particular, high-energy protons, propagate along the disk axis and interact with photons from the disk and produce high-energy neutrinos. 

It is known from a number of observations that AGN are also bright in the X-ray range \cite{Ajello:2008xb, Fabian:1992rt, Ueda:2014tma, Hasinger:2005sb}. One of the possible mechanisms of formation is the presence of the hot component (so-called corona) of electrons with energies $\sim$ 100 keV. Low-energy photons of the disk undergo inverse Compton scattering on the hot electrons, and as a result their frequency increases to X-ray range.

\subsubsection*{Geometry}

Despite many studies in theory and X-ray observations, the question of nature and geometry of the AGN corona is still open. Different heating mechanisms propose different geometries (see e.g. review \cite{Karas:2019lju}). However, Recent X-ray measurements in Seyfert galaxies favor an extended geometry of the corona rather than a point-like geometry \cite{Gianolli:2023zji, Tagliacozzo:2023apk, Ingram:2023twz}. In this work, we consider a simple two-phase model. There, the cold phase is represented by a geometrically thin Shakura-Sunyaev accretion disc \cite{Shakura:1972te}.  A part of it is located in a hot plane-parallel cloud, where the temperature inside is radially averaged due to Comptonization. The remaining part, which extends beyond the cloud, behaves as a standard accretion disk. More detailed two-component models are presented, for example, in \cite{Haardt:1991tp, Liu:2002ts}.

Various observations point to the compactness of the X-ray corona. In line with microlensing observations \cite{Chartas:2015mta} and spectral analyzes \cite{Jin:2011wn}  the coronal size is typically scaled as $R_c\sim 30 r_g$ for all SMBH, where $r_g$ is gravitational radius,
\begin{equation}
    r_g=3\cdot10^{13} \left( \frac{M}{10^8 M_{\odot}} \right) \ \text{cm},
\end{equation}
$M$ is the SMBH mass.

\subsubsection*{Spectrum}
To find the spectrum of the two-component disk-corona system, it is necessary to solve the Kompaneets equation \cite{Kompaneets:1957}, which describes the Comptonization of radiation on electrons. Assuming that the source of low-energy photons is located inside the electron plasma cloud with electron temperature $T_e$, the solution can be written as \cite{Sunyaev:1980fq}:
\begin{equation}
    F(x)=\int_{0}^{\infty} \frac{1}{x_0}G(x,x_0)f(x_0)dx_0,
    \label{spec Sun}
\end{equation}
where $x=p/T_e$, is dimensionless photon energy $p$ (we assume $\hbar = c = 1$), $f(x)$ is the spectrum of low-frequency photons, $G(x,x_0)$ is the Green function given in \cite{Sunyaev:1980fq} which contains information about Thomson optical depth $\tau_\text{T}$ and temperature $T_e$:

\begin{equation}
     G(x,x_0) = \frac{\alpha(\alpha+3)}{2\alpha+3} \left( \frac{x}{x_0} \right)^{3+\alpha}, \ \ 0<x<x_0
\end{equation}
and for $x\ge x_0$
\begin{eqnarray}
G(x,x_0) = \frac{\alpha(\alpha+3)}{\Gamma(2\alpha+4)} \left( \frac{x_0}{x} \right)^{\alpha} \exp(-x) 
\times \int_0^{\infty} t^{\alpha-1} \exp(-t)(x+t)^{\alpha + 3} dt,
\end{eqnarray}
where 
\begin{equation}
    \alpha = \left(\frac{9}{4} +\frac{\pi^2m_e}{3T_e(\tau_{\text{T}} +2/3)^2}  \right)^{1/2} -\frac{3}{2},
    \label{spectral_index}
\end{equation}
$m_e$ is the electron mass and $\Gamma(x)$ is the gamma function.

Note, that the spectral index $\alpha$ given in (\ref{spectral_index}) is correct for $\tau_T \geq 3$ and $T_e \ll m_e$\cite{Sunyaev:1980fq}. For weakly relativistic plasma ($T_e \approx 50-100$ keV) and for $\tau_T \leq 1$ keV can be used good approximation \cite{Pozd:1979ast}:
\begin{equation}
    \alpha = \frac{-\lg \tau_T + 2/(n+3)}{\lg(12n^2 + 25n)},
    \label{Alp_Pozd}
\end{equation}
where $n = T_e/m_e$. For more details see review \cite{Pozd:1983sun}

The standard model of an accretion disk is the multitemperature Shakura–Sunyaev disk with radial dependence $\displaystyle \large T(r) \propto T_d \, r^{-3/4}$, where $T_d$ is the effective temparature \cite{Shakura:1972te} . However, all existing treatments of Comptonization problem assume the soft photon source to have a single-temperature (isothermal) blackbody spectrum \cite{Sunyaev:1980fq, Pozd:1983sun, Ueda:2014tma, Zdziarski:2019cvs}. To reconcile these approaches, for the portion of the disk situated within the coronal region, we employ a radially averaged temperature $T_{\text{av}}$. It can be shown that $T_{\text{av}} \sim 0.1-0.2 \, T_d$ on scales of order $10r_g$. For the outer part of the disk, which lies outside the corona, we adopt the standard radial temperature dependence \cite{Shakura:1972te}.

Therefore, for the function $f(x)$ in Eq. \ref{spec Sun}, we adopt a thermal spectrum with a temperature $T_{\text{av}}$. This average temperature is derived by radially averaging the effective temperature profile of the multitemperature disk and is expressed in terms of the disk's effective temperature $T_d$.

\begin{equation}
    f(x) = \frac{a^3}{2\zeta(3)} \frac{x^3}{\exp(ax)-1},
    \label{bb spec}
\end{equation}
$a=T_e/T_{\text{av}}$, $\zeta(x)$ is the Dzeta-Rieman function. Therefore, from (\ref{spec Sun}) and (\ref{bb spec}), the photon spectral number density after Comptonization on a corona surface reads
\begin{equation}
     n(p, \Omega) =2\frac{T_e^2}{\pi} \int_{0}^{\infty} \frac{1}{x} G(x,x_0) \frac{x_0^2}{\exp(ax_0)-1} dx_0.
\end{equation}

Spectrum (\ref{spec Sun}) after substituting the blackbody spectrum into it in the low energy region preserves the Rayleigh-Jeans asymptotics \cite{Pozd:1983sun, Pozd:1979ast, Sunyaev:1980fq}. In a wide energy region the spectrum is described by a power law with spectral index $\alpha$ with high energy cut \cite{Pozd:1979ast}. From spectral analyses of the AGN, the spectral index for X-ray radiation is considered to be equal in order of 1.9 (for $dN/dE$) \cite{Ueda:2003yx, Ueda:2014tma}. For $T_e=100$ keV and $\tau_T=1$, the spectral index $1+\alpha$ is 1.88 from eq. (\ref{Alp_Pozd})

\begin{figure}
    \centering
    \includegraphics[width=0.8\textwidth]{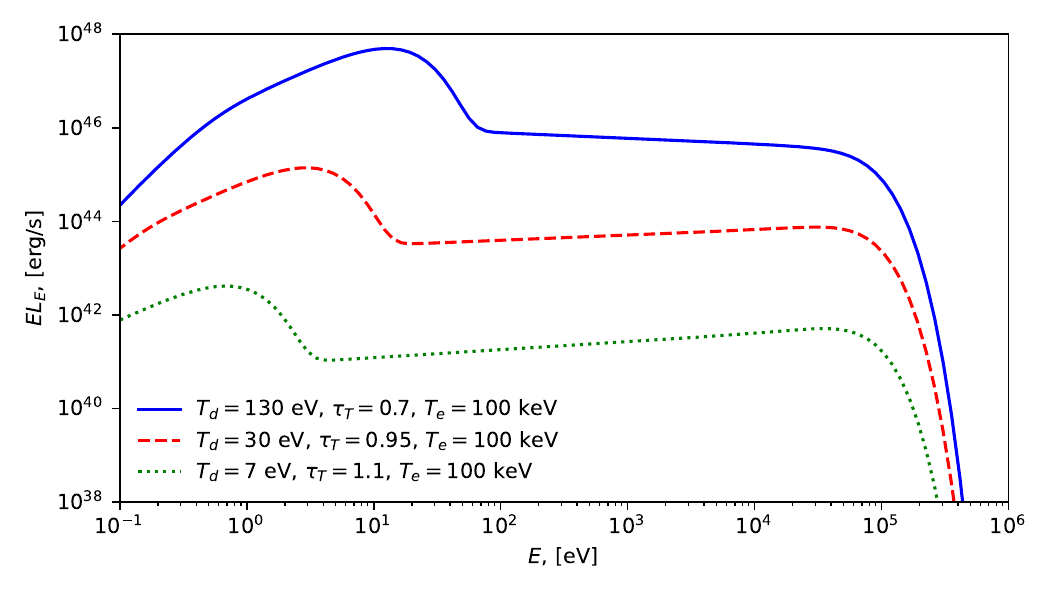}
    \caption{SEDs arising from an AGN with a coronal size of 15$r_g$ for different luminosities $10^{42} \text{ erg/s}, \, 10^{44}$ erg/s and $10^{46}$ erg/s from bottom to top.}
    \label{sed}
\end{figure}

As an example, Fig. \ref{sed} shows the rest frame spectral energy distribution (SED) for an AGN arising from the accretion disk and corona for some parameters.

\subsubsection*{Proton acceleration and propagation}
The mechanism of particle acceleration in the corona remains highly ambiguous. Examples of various acceleration scenarios can be found, for instance, in \cite{Levinson, Neronov:2009zz, Blandford:1987pw, Zhdankin:2018lhq, Hoshino:2013pza}. Since we are primarily interested in the neutrino spectrum, for the sake of simplicity, we restrict ourselves to the scenario outlined in \cite{Kalashev:2014vya}. Following \cite{Kalashev:2014vya}, we assume that protons are accelerated by electric fields \red in the magnetosphere in the close vicinity of the SMBH \black and then released at some point $z_0$ from it. The injected proton spectrum at point $z_0$ is assumed to be proportional to $E^{-1}$, with a high-energy cut $E_{\text{max}}$. The proton injection spectrum is motivated by recent Particle-in-Cell simulations at the vicinity of SMBH \cite{Kisaka:2020lfl}. As well as in \cite{Kalashev:2014vya} we assume $z_0=2r_g$ We focus on accelerated protons with an initial energy $\sim$ 10-100 PeV, which corresponds to a threshold photon energy of 1-10 eV.

After injection, protons propagate near the vicinity of the SMBH along the disk axis and interact with the disc-corona photons. High-energy neutrinos are produced in $p\gamma$ reactions and freely escape from the AGN. 

\begin{figure*}[t!]
    \centering
    \includegraphics[width=1.1\textwidth]{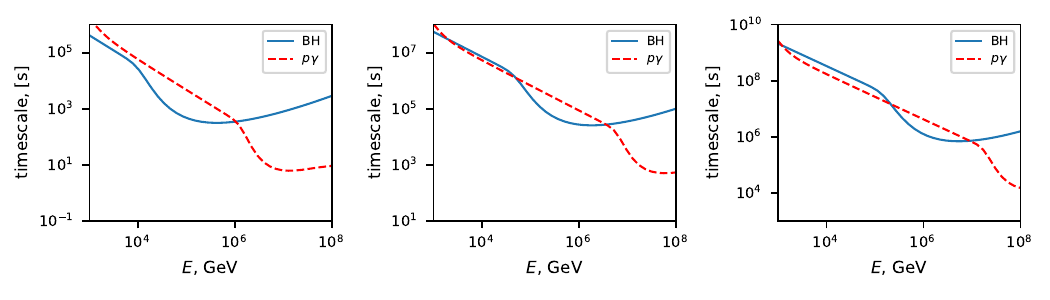}
    \caption{Comparisons of timescales of energy losses as a function of proton energy. The red dashed curve corresponds to photohadronic losses, while the solid blue curve represents Bethe-Heitler processes. The timescales are presented for various $L_X = 10^{46}$ erg/s (left), $10^{44}$ erg/s (middle), $10^{42}$ erg/s (right). The parameters are same as on Fig.\ref{sed}.}
    \label{timescales}
\end{figure*}

Within our model, neutrinos are produced through photohadronic interactions. However, Bethe-Heitler processes ($p\gamma\to e^+e^-p$) may dominate at lower energies. To determine the energy range where this transition occurs, the energy-loss timescale is calculated for both reaction types:

\begin{eqnarray}
    t^{-1} = \int d\Omega \int dE \, (1-\cos \theta) n(\mathbf{p}, r, z) \sigma(\epsilon_r) K(\epsilon_r), 
\end{eqnarray}
where $K(\epsilon_r)$ is the inelasticity of the collision. The results are shown in Fig.\ref{timescales} for various X-ray luminosities $L_X$.

\red
In this work, we neglect turbulence-related effects for two reasons. First, particle acceleration occurs in the magnetosphere, in the immediate vicinity of the black hole, where strong, ordered electromagnetic fields dominate the acceleration process. Second, protons interact with photons near the SMBH with very high efficiency.
\black

Indeed, the spatial diffusion mean free path $\lambda$ (related to the diffusion coefficient by $D = \lambda c/3$) is given by $\lambda = \eta_g E/(e B)$, where $\eta_g$ is the gyrofactor ($\eta_g = 1$ corresponds to the Bohm limit), $B$ the magnetic field, $e$ the proton charge, and $E$ the proton energy. \red For our estimates, we adopt $B = 10^3$ G as a plausible value for this acceleration model \cite{Ptitsyna:2015nta, Kisaka:2020lfl}. The gyrofactor can span a wide range of values; for example, in blazars $\eta_g$ can be as high as $\sim 10^{4}$ \cite{Finke:2008pe}, and in some individual sources it can reach $10^{5}$ \cite{Inoue:2016fwn}. Meanwhile, \cite{Inoue:2019fil} suggests a favored value of $\eta_g \sim 30$ for Seyfert galaxies, obtained under the assumption of a weak coronal magnetic field ($\sim 10$~G) and diffusive shock acceleration. By contrast, the acceleration mechanism considered in this paper implies substantially larger values of the gyrofactor. Adopting values of $\eta_g \sim 10^3 - 10^5$ for the estimate, $\lambda$ in units of $r_g$ can be in the range $1 r_g-10^2 r_g$ at a typical energy of 10 PeV, where $p\gamma$ interactions dominate over Bethe–Heitler processes.
\black

The mean free path for $p\gamma$ interactions, $\lambda_{p\gamma}$, can be calculated using the relation 
\begin{equation}
    \displaystyle 1 = \int_{z_0}^{\lambda_{p\gamma}} R(z, E)  dz \ ,    
\end{equation}
where $R(z,E)$ is defined below (Eq. \ref{rate3}). Within the framework of the model adopted in this work, $\lambda_{p\gamma} \sim 10^{-3} r_g$, $10^{-1} r_g$, and $10^{2} r_g$ for luminosities $L_X = 10^{46}$, $10^{44}$, and $10^{42}$ erg/s, respectively and $E = 10$ PeV. The parameter values corresponding to these luminosities are listed in Figure \ref{sed}. Thus, turbulence significantly affects the outcome for sufficiently faint galaxies, namely those with $L_X \leq 10^{42}$ erg/s. We note that turbulent acceleration becomes effective at $L_X \sim 10^{42}$ erg/s and below, a result previously reported in \cite{Murase:2019vdl}.

The timescale for turbulent acceleration can be estimated as follows:

\begin{eqnarray}
    t_{T} = \frac{1}{\zeta} \left( \frac{v_{\mathrm{A}}}{c} \right)^{-2} \frac{R_c}{c} \left( \frac{r_{\mathrm{L}}}{R_c} \right)^{2-q} \gamma^{2-q},
\end{eqnarray}
here $\zeta$, represents the turbulent-to-background magnetic field strength ratio, $v_{\mathrm{A}}$ is the Alfvén speed, $r_{\mathrm{L}}$ - is the Larmor radius.
\red
Using $B=10^3$ G, and assuming the Kolomogorov spectrum ($q=5/3$) the estimate can be rewritten in the form:

\black

\begin{eqnarray}
   \displaystyle t_T \simeq 1.5 \times 10^4 \ \red \frac{1}{\zeta} \black \  \left(\frac{\tau_{\mathrm{T}}}{1.1} \right)  \left( \frac{R_c}{10r_g} \right)^{-1/3}  \left( \frac{M_{{}_{\mathrm{BH}}}}{10^8 M_{\odot}} \right)^{-1/3}  \left( \frac{B}{\red 10^3 \, \mathrm{G} \black} \right)^{-7/3}  \left(\frac{E}{\red 1\text{ PeV} \black} \right)^{1/3}   [\text{ s }].
   \label{turbo}
\end{eqnarray}

\red
In models of stochastic acceleration, it is typically assumed that $\zeta < 1$ (except in weakly magnetized plasmas; see \cite{Kimura:2018clk} and references therein). For instance, \cite{Murase:2019vdl} adopts $\zeta \sim 0.1$. However, these models do not directly apply to our scenario, which involves particle acceleration in a strongly magnetized magnetosphere near the black hole horizon. In such a highly magnetized environment, we expect $\zeta$ to be significantly smaller, $\zeta \ll 1$. Therefore, turbulent acceleration remains inefficient compared to the $p\gamma$ cooling rate across the neutrino production energy range ($\ge 1$--$10$~PeV).  
\black

\red Thus, based on the above estimates, neutrinos are produced predominantly near the black hole in the magnetospheric acceleration scenario considered here. \black The secondary photons, electrons, and positrons interact with the disk+corona radiation field. However, due to the large optical depth (see the Appendix for details), the gamma-ray radiation is absorbed within the source, consistent with the "hidden source" scenario for neutrino production \cite{Murase:2015xka} and avoiding an overproduction of the Fermi-LAT observed gamma-ray flux \cite{Fermi-LAT:2014ryh}.

\section{Calculation}
\label{3}

In this section, we outline the scheme for simulating proton propagation in the manner proposed in \cite{Kalashev:2014vya}, where protons are assumed to propagate along the disk axis. This is a simplification since protons may be ejected at different angles due to processes occurring during their acceleration. The simulation scheme can be generalized to an arbitrary propagation direction (see Appendix). However, having performed simulations for various directions, we found that the resulting neutrino spectrum is practically insensitive to these variations for the flat geometry used in this work. For models that consider the detailed spatial structure of the corona, this approximation may prove inadequate.

\subsubsection*{Preliminary}
The photon density created by the small segment of the corona + disk system $rdrd\varphi$ at distance $r$ from the the SMBH and distance $z$ along the disk axis:

\begin{equation}
n(\mathbf{p},r,z) = \frac{1}{2\pi} \frac{\delta^{(3)} (\mathbf{n}-\mathbf{n_{0}}) rdrd\varphi}{z^2+r^2}n(p, \Omega),
\end{equation}
where $\mathbf{n_0}$ is the unit vector in the direction from the small corona or disk segment to $z$ and $\mathbf{p}$ - photon momentum. Because of the model is cylindrical symmetry the integration over $\varphi$ gives an additional factor of $2\pi$. Contribution from this segment to the reaction rate for $p\gamma$ interaction $R(E,z,r)$ is:
\begin{equation}
    R(E,z,r) = \int d^3 \mathbf{p} (1- \beta \cos\theta) n(\mathbf{p},z,r) \sigma(\epsilon_r),
    \label{Rate1}
\end{equation}
where $E$ is the proton energy, $\beta=1/\sqrt{1-\gamma^2}$, where $\gamma$ is the proton gamma factor, $\cos\theta$ is cosine of the angle between proton and photon momenta, $\sigma(\epsilon_r)$ is total cross section of the photohadronic interaction, $\epsilon_r=p\gamma(1-\beta\cos\theta)$ is photon energy in the proton rest frame.

In the case of ultra-relativistic protons after integration (\ref{Rate1}) over the angular part, one obtains

\begin{equation}
    R(E,z,r) = \frac{1-\cos\theta}{z^2+r^2} \int dp p^2 n(p,z,r) \sigma(\epsilon_r),
    \label{rate2}
\end{equation}
where $\cos\theta = z/\sqrt{z^2+r^2}$.

Reaction rate at point $z$:

\begin{equation}
    R(E,z) = \int rdr R(E,z,r).
    \label{rate3}
\end{equation}

The proton optical $\tau_p$ depth can be calculated using the formula:

\begin{eqnarray}
    \tau_p = \int_{z_0}^{\infty} R(E,z) dz.
    \label{opt_depth_proton}
\end{eqnarray}
 Typical values of $\tau_p \sim 10^4-10^5$ for a proton energy on the order of PeV.

\subsubsection*{Simulation}
To calculate the spectrum from one source, we use the Monte Carlo approach as proposed in \cite{Kalashev:2014vya}. We briefly outline this method here.
 
Firstly, we need to simulate proton propagation along the disk axis. For this, in each iteration random numbers $\xi_1, \xi_2, \xi_3$ distributed in [0,1] are selected. The travelled optical depth $\tau_j$ at $j$-th iteration is sampled,

\begin{equation}
    \tau_j = -\log \xi_1 ,
\end{equation}
A point of $j$-th interaction $z_j$ is obtained by the resolution of the integral equation:
\begin{equation}
    \tau_j = \int_{z_{j-1}}^{z_j} R(E,z) dz.
\end{equation}
If the equation has no solution, it means that the nucleon escape from the AGN. 

The interaction angle $\cos \theta_j = z_j/\sqrt{z_j^2 + r_j^2}$ is sampled using (\ref{rate2}), (\ref{rate3}):
\begin{equation}
    \xi_2 = \frac{1}{R(E,z_j)}\int_{r_g}^{r_j} rdr R(E,z_j,r).
\end{equation}
Momentum of interacting photon $p_j$ is modelled similarly:
\begin{equation}
    \xi_3 = \frac{1}{R(E,z_j,r_j)} \frac{1-\cos\theta_j}{z_j^2+r_j^2} \int_0^{p_j} dp p^2 n(p,z_j,r_j) \sigma(\epsilon_r).
\end{equation}
Then the SOPHIA event generator \cite{Mucke:1999yb} is used to simulate $p\gamma$ reactions until the nucleons (and antinucleons) escape the interaction region or until an energy is reached at which Bethe–Heitler processes dominate over $p\gamma$ interactions. As a result, after propagation of large number of protons we get the list of secondary particles and their momenta.

Finally, we integrate over the sources taking into account their cosmological evolution:

\begin{eqnarray}
 \Phi_{\nu}(E) = \frac{c}{4\pi H_0} \int dz \frac{1}{\sqrt{\Omega_M (1+z)^3 + \Omega_{\Lambda}}}  \int d\log L_X \frac{d\rho (z, L_X)}{d\log L_X} \frac{L_{\nu}(E', L_X)}{E'},
 \label{evolution}
\end{eqnarray}
where $d\rho/d\log L_X$ is the X-ray luminosity function, $L_{\nu}$ is neutrino luminosity per X-ray luminosity $L_X$ per energy $E' = (1+z)E$, $H_0$ is the local Hubble constant, $\Omega_M=0.3, \ \Omega_{\Lambda}=0.7$. As a X-ray luminosity function we take the expression obtained in \cite{Ueda:2014tma} (see eq. (16)-(19) and Table 4 of Ref. \cite{Ueda:2014tma}). We use the following approximation: that for fixed parameters $T_e$, $T_d$ and $\tau_T$, the change in $L_X$ is associated with a change in the size of the corona $R_c$. Therefore, the neutrino luminosity is assumed to be directly proportional to X-ray luminosity $L_{\nu} \propto L_X$.

\begin{figure}
    \centering
    \includegraphics[width=0.7\textwidth]{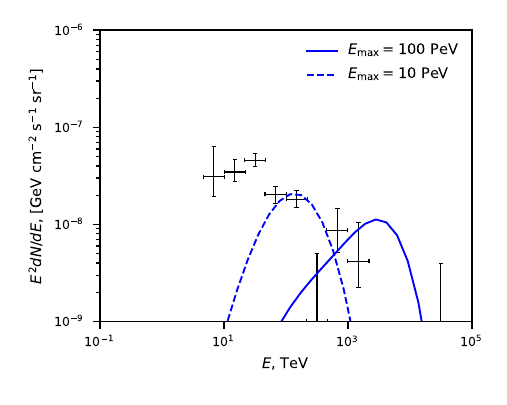}
    \caption{Resulting diffuse neutrino flux from AGN with cosmological evolution \ref{evolution} for $T_e = 100$ keV, $T_d = 130$ eV (which corresponds $T_{\text{av}} = 15$ eV), $\tau_\text{T}=0.65$ for various $E_{\text{max}}=10$ and 100 PeV. The black data points is the IceCube-measured spectrum of the astrophysical neutrino background \cite{Naab:2023xcz}.} 
    \label{100kev}
\end{figure}

After that we normalize the resulting neutrino flux to the IceCube diffuse flux. The normalization factor corresponds to the ratio of neutrino and X-ray luminosities: $L_{\nu} = 0.08 L_X$. The results are shown in Fig. \ref{100kev}.

\section{Discussion}
\label{4}

The observed astrophysical neutrino flux has different contributions, and only a part of it may be associated with AGN - e.g. the Galactic contribution is presently well established, but the statistics is insufficient to firmly separate the flux into different components. However, observations of $\sim$ 10-TeV neutrinos from AGN by IceCube implies that the accretion-disk thermal radiation cannot serve as an efficient target for neutrino production at these energies \cite{Kalashev:2014vya, Murase:2019vdl, Neronov:2025cfc}. The present work demonstrates that the account of the corona emission provides conditions for efficient neutrino production in a $\sim 50$ TeV - sub PeV energy band. Given the presence of other contribution, we do not attempt to fit the observed total neutrino spectrum, and demonstrate instead the order-of-magnitude matching of the fluxes for reasonable values of the parameters.

 In total, the model has six parameters, $T_e$, $T_d$, $\tau_T$, $E_\text{max}$, $z_0$ and the proton spectral index. We choose natural values of the parameters. The proton spectral index is assumed to be 1, which is motivated by Particle-in-Cell simulations  \cite{Kisaka:2020lfl}. Following previous work \cite{Kalashev:2014vya}, we assume that the acceleration occurs in the black hole magnetosphere, which sets $z_0 = 2r_g$. Due to the large optical depth for protons, within small variations, this does not greatly affect the neutrino spectrum in the TeV-PeV region. The values of $T_e$ and $\tau_T$ are chosen in accordance with the spectral index for AGN obtained in \cite{Ueda:2003yx}. The temperature of the accretion disk $T_d$ is chosen to approximately match the SED of an AGN.
The model dependence on $E_{\text{max}}$ is shown in Fig. \ref{100kev} for $E_{\text{max}}$ = 10 PeV and 100 PeV.  The model dependence on $T_d$ was studied in \cite{Kalashev:2014vya} but for different proton spectral index. 
 
 The  proportionality coefficient corresponds $L_{\nu} = 0.08 L_X$, which is a reasonable value, roughly corresponds to the overall normalization of the integral flux of all sources on the IceCube diffuse neutrino flux. We demonstrate by population analysis that a reasonable neutrino luminosity of an AGN does not overproduce neutrinos with respect to the IceCube observations. Moreover, the model predicts a suppression of gamma-ray emission; in other words, the neutrino sources could be "hidden" AGN cores \cite{Murase:2015xka}. 

\begin{figure}
    \centering
    \includegraphics[width=0.7\textwidth]{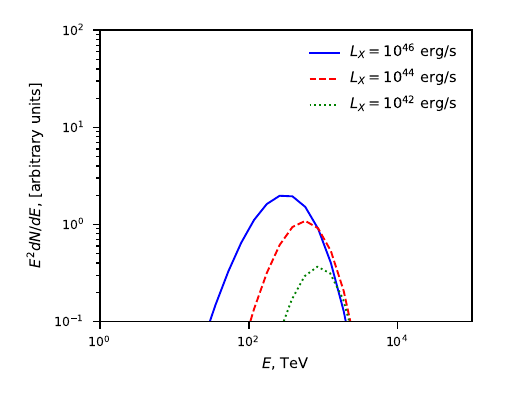}
    \caption{The non-normalized neutrino flux from a single AGN source for various $L_X$. The parameters are same as on Fig.\ref{sed} and $E_{\text{max}} = 10$ PeV.}
    \label{models}
\end{figure}

\
Figure \ref{models} shows how the X-ray luminosity affects the shape of the neutrino spectrum from a single source. As $L_X$ increases, the neutrino spectrum broadens because the threshold for photohadronic interactions shifts to lower energies at higher luminosities.

Each individual AGN has its own set of parameter values, and all of them contribute to the neutrino spectrum. Our qualitative analysis focuses on demonstrating that different AGN (i.e., AGN with different parameters within naturally motivated physical ranges) could potentially explain both the $\sim$ 50 TeV and sub PeV energy ranges simultaneously. However, the neutrino spectrum can extend to lower energies if proton-proton interactions are taken into account \cite{Inoue:2019fil, Murase:2019vdl}.

The calculations were carried out assuming proton propagation along the disk axis, a convenient choice owing to the azimuthal symmetry it provides. The computational technique can be generalized (see Appendix) to scenarios where protons propagate at a nonzero angle relative to the disk axis. However, simulations of the neutrino spectrum at different angles showed that the spectrum is insensitive to these variations within the framework of the flat corona model. Consequently, the adopted proton propagation direction does not influence the result while offering the benefit of computational efficiency. In order to illustrate the weak dependence on the propagation angle, Figure \ref{time_chi} shows the calculated interaction timescales for different angles, corresponding to $0$, $\pi/6$, and $\pi/3$. The effect might become appreciable if the detailed internal structure of the corona were considered; however, this lies well beyond the scope of the present paper, and we defer this matter to future investigations.

It is also useful to make comparisons with other models. A similar calculation, but only for an anisotropic accretion disk with, was made by Kalashev et al. \cite{Kalashev:2014vya}. In \cite{Kalashev:2014vya} the spectrum is narrower because the accretion disk photons have an energy $\sim 10-100$ eV and therefore this model can explain narrow bumps in the observed spectrum. However it is also worth noting that in \cite{Kalashev:2014vya}, the spectral index of the injected protons is assumed to be 2, which shifts the neutrino spectrum to lower energies. In \cite{Murase:2019vdl} and \cite{Inoue:2019fil} the phenomenological disk-corona models based on the observational spectra of AGN and empirical relations are presented. These papers take magnetic fields into account and consider specific acceleration models.

We focus on the propagation of protons in the central regions of AGN using the Monte Carlo technique, using only the spectrum of the corona and assuming acceleration in the magnetosphere in the vicinity of the SMBH.

\section{Conclusions}

In this paper, the model proposed in \cite{Kalashev:2014vya} was generalized to the case of the presence of the hot corona. It is capable of explaining both $\sim 50$ TeV-PeV neutrinos in IceCube within one and the same production mechanism and class of sources. The simulation scheme has also been generalized to proton propagation at an arbitrary angle relative to the accretion disk axis. Within the framework of the flat geometry, no significant dependence of the neutrino spectrum on the proton ejection direction was found. Under an alternative assumption regarding the acceleration and propagation mechanism, our results remain consistent with other models. \cite{Inoue:2019fil, Murase:2019vdl}.

Further development of the model includes taking into account magnetic fields, modeling the propagation of secondary particles in AGN and taking into account the detailed structure of the corona.


\acknowledgments

The author is grateful to Sergey Troitsky and Oleg Kalashev for valuable remarks throughout the entire work and for careful reading of the manuscript and Elena Seifina for useful discussion. This work is supported in the framework of the State project “Science” by the Ministry of Science and Higher Education of the Russian Federation under the contract 075-15-2024-541. 
The author also thanks the Theoretical Physics and Mathematics Advancement Foundation “BASIS” for the fellowship under the contract 24-2-1-101-1.

\appendix

\section{Appendix: Angular dependence}
In Section \ref{3}, the proton propagation occurred along the disk axis. This direction has an advantage from the computational perspective, since in this case there is azimuthal symmetry, which allows integration over the azimuthal angle $\varphi$. In the general case, the direction of proton motion may be at a certain angle $\chi$. Here we demonstrate how the equations are transformed in the case of a nonzero $\chi$.

Let $z$ denote the distance from the proton to the black hole along the propagation direction. Then, the contribution to the reaction rate originating from the segment $r dr d\varphi$ is modified (cf. \ref{rate2}):

\begin{equation}
    R(E,z,r, \varphi) = \frac{1-\cos\theta}{z^2+r^2 - 2zr\sin\chi \cos \varphi} \int dp p^2 n(p,z,r) \sigma(\epsilon_r),
    \label{rate2_appendix}
\end{equation}
where $\cos\theta = z \cos\chi/\sqrt{z^2+r^2 - 2zr\sin\chi\cos\varphi}$.

Consequently, a dependence on the azimuthal angle $\varphi$ emerges, which must also be accounted for in Monte Carlo simulations. Note that this expression reduces to (\ref{rate2}) in the limit $\chi \to 0$. The simulation of proton propagation is performed in the same manner as in Section \ref{3}, except that $\varphi$ also needs to be simulated at each $j$-th interaction:

\begin{equation}
    \displaystyle \xi = \frac{\displaystyle \int_{0}^{\varphi_j} d\varphi R(E,z_j,r_j, \varphi)}{\displaystyle \int_{0}^{2\pi} d\varphi R(E,z_j,r_j, \varphi)},
\end{equation}
where $\xi$ - random number unifromly distributed in [0,1].

Calculations within the framework of the described flat corona model show that the neutrino spectrum is insensitive to changes in the angle $\chi$. For illustration, we show a plot of the interaction timescales for various $\chi$.

\begin{figure}
    \centering
    \includegraphics[width=0.8\textwidth]{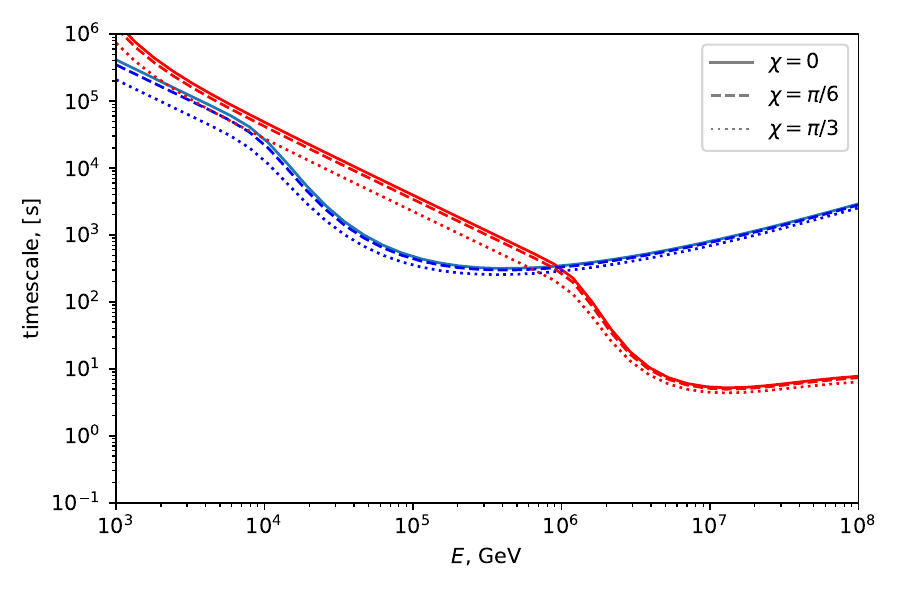}
    \caption{ Dependence of interaction timescales on the angle $\chi$. The blue and red curves represent the Bethe-Heitler process and $p\gamma$ interactions, respectively. Line styles denote different angles: solid ($\chi = 0$), dashed ($\chi = \pi/6$), and dotted ($\chi = \pi/3$). The curves are computed for $T_e = 100$ keV, $T_d = 130$ eV, and $\tau_T = 0.7$.} 
    \label{time_chi}
\end{figure}

\section{Appendix: Electromagnetic cascades}

For the sake of completeness, we provide here a brief discussion on the calculation of electromagnetic cascades. To estimate the spectrum of secondary photons and electrons, one can solve the one-dimensional stationary transport equations. Transport equations incorporating inverse Compton scattering and photon-photon annihilation can be written in the following form for electrons (or positrons):

\begin{eqnarray}
    c\frac{\partial n_e}{\partial z} = -n_e(p)\int n_{\text{ph}}(\epsilon)(1-\mu) \frac{d\sigma_{\text{ICS}}}{dp'}\, d\epsilon\, d\mu \, dp' + \nonumber \\[2mm] \int n_e(p') n_{\text{ph}}(\epsilon)(1-\mu) \frac{d\sigma_{\text{ICS}}}{dp}\, d\epsilon\, d\mu \, dp' + \nonumber \\[2mm]
    \frac{1}{2}\int n_{\gamma}(p') n_{\text{ph}}(\epsilon)(1-\mu) \frac{d\sigma_{\gamma \gamma}}{dp}\, dp'\, d\mu \, d\epsilon
\end{eqnarray}

and for photons:

\begin{eqnarray}
    c\frac{\partial n_{\gamma}}{\partial z} = -n_{\gamma}(p)\int n_{\text{ph}}(\epsilon)(1-\mu) \frac{d\sigma_{\gamma \gamma}}{dp}\, d\epsilon\, d\mu \, dp_1 +\nonumber \\[2mm] 
    \int n_e(p') n_{\text{ph}}(\epsilon)(1-\mu) \frac{d\sigma_{\text{ICS}}}{d p}\, d\epsilon\, d\mu \, dp' .
\end{eqnarray}
Where $n_{\text{ph}} , \, n_e ,\, n_{\gamma}$ - are the spectral number density of the background photons, secondary electrons and secondary photons, $z$ - is distance along disk axis and $d\sigma_{\text{ICS}}, \, d\sigma_{\gamma \gamma}$ - are differential cross sections for inverse Compton scattering and $e^{+} e^{-}$ production.

It proves useful to linearize these equations using the substitution $\displaystyle n^i = \int_{p_{i-1/2}} ^ {p_{i+1/2}} n(p) \, dp$, where the energy is binned, and the central value of the $i$-th bin is labeled $p_i$ while its boundaries are $p_{i-1/2}, \, p_{i+1/2}$. The equations can then be rewritten in the following form \cite{Lee:1996fp, Kalashev:2014xna}:

\begin{eqnarray}
    c\frac{\partial n_{e}^i}{\partial z} = -n_e^i A_{e,i} + \sum_j n_e^j B_{i,j} + \sum_j n_{\gamma}^j C_{j,i},
\end{eqnarray}
where

\begin{eqnarray}
    A_{e,i} = \int n_{\text{ph}}(\epsilon)(1-\mu) \frac{d\sigma_{\text{ICS}}(p_i, \epsilon, p')}{dp'}\, d\epsilon\, d\mu \, dp' , \\[2mm]
    B_{i,j} = \int d\epsilon \, n_{\text{ph}}(\epsilon)(1-\mu) \int_{p_{i-1/2}} ^ {p_{i+1/2}} \frac{d\sigma_{\text{ICS}}(p_j, \epsilon, p)}{dp}\, \, d\mu \, dp  \nonumber \\[2mm]
     C_{i,j} = \int d\epsilon \, n_{\text{ph}}(\epsilon)(1-\mu) \int_{p_{i-1/2}} ^ {p_{i+1/2}} \frac{d\sigma_{\gamma \gamma}(p_j, \epsilon, p)}{dp}\, \, d\mu \, dp  \nonumber \\[2mm]
\end{eqnarray}

Numerical analysis has shown that this system of differential equations is highly stiff, making it difficult to solve without introducing additional simplifications. The term with coefficients $A_i$ dominates on the right-hand side. From a physical standpoint, this indicates a large optical depth for secondary particles in the energy range starting from 10 GeV. Therefore, although the calculation is computationally expensive, it can be argued that in this model, the gamma-ray flux will be suppressed. Thus, the model also predicts a "hidden" neutrino source \cite{Murase:2015xka}.

\bibliographystyle{JHEP}
\bibliography{biblio.bib}

@article{Kalashev:2014vya,
    author = "Kalashev, O. and Semikoz, D. and Tkachev, I.",
    title = "{Neutrinos in IceCube from active galactic nuclei}",
    eprint = "1410.8124",
    archivePrefix = "arXiv",
    primaryClass = "astro-ph.HE",
    reportNumber = "INR-TH-2014-025",
    doi = "10.1134/S106377611503022X",
    journal = "J. Exp. Theor. Phys.",
    volume = "120",
    number = "3",
    pages = "541--548",
    year = "2015"
}

@article{Chartas:2015mta,
    author = "Chartas, G. and Rhea, C. and Kochanek, C. and Dai, X. and Morgan, C. and Blackburne, J. and Chen, B. and Mosquera, A. and MacLeod, C.",
    title = "{Gravitational Lensing Size Scales for Quasars}",
    eprint = "1509.05375",
    archivePrefix = "arXiv",
    primaryClass = "astro-ph.HE",
    doi = "10.1002/asna.201612313",
    journal = "Astron. Nachr.",
    volume = "337",
    number = "4/5",
    pages = "356--361",
    year = "2017"
}

@article{Haardt:1991tp,
    author = "Haardt, F. and Maraschi, L. af Milan U.",
    title = "{A two-phase model for the X-ray emission from Seyfert galaxies}",
    doi = "10.1086/186171",
    journal = "Astrophys. J. Lett.",
    volume = "380",
    pages = "L51--L54",
    year = "1991"
}

@article{Liu:2002ts,
    author = "Liu, B. F. and Mineshige, S. and Meyer, F. and Meyer-Hofmeister, E. and Kawaguchi, T.",
    title = "{Two-temperature coronal flow above a thin disk}",
    eprint = "astro-ph/0204174",
    archivePrefix = "arXiv",
    doi = "10.1086/341138",
    journal = "Astrophys. J.",
    volume = "575",
    pages = "117",
    year = "2002"
}

@article{Jin:2011wn,
    author = "Jin, C. and Ward, M. and Done, C. and Gelbord, J. M.",
    title = "{A Combined Optical and X-ray Study of Unobscured Type 1 AGN. I. Optical Spectra and SED Modeling}",
    eprint = "1109.2069",
    archivePrefix = "arXiv",
    primaryClass = "astro-ph.HE",
    doi = "10.1111/j.1365-2966.2011.19805.x",
    journal = "Mon. Not. Roy. Astron. Soc.",
    volume = "420",
    number = "3",
    pages = "1825--1847",
    year = "2012"
}

@article{Sunyaev:1980fq,
    author = "Sunyaev, R. A. and Titarchuk, L. G.",
    title = "{Comptonization of X-rays in plasma clouds. Typical radiation spectra}",
    journal = "Astron. Astrophys.",
    volume = "86",
    pages = "121--138",
    year = "1980"
}

@article{Mucke:1999yb,
    author = "Mucke, A. and Engel, Ralph and Rachen, J. P. and Protheroe, R. J. and Stanev, Todor",
    title = "{SOPHIA: Monte Carlo simulations of photohadronic processes in astrophysics}",
    eprint = "astro-ph/9903478",
    archivePrefix = "arXiv",
    reportNumber = "BA-99-33, ADP-AT-99-3",
    doi = "10.1016/S0010-4655(99)00446-4",
    journal = "Comput. Phys. Commun.",
    volume = "124",
    pages = "290--314",
    year = "2000"
}

@article{Kalashev:2014xna,
    author = "Kalashev, O. E. and Kido, E.",
    title = "{Simulations of Ultra High Energy Cosmic Rays propagation}",
    eprint = "1406.0735",
    archivePrefix = "arXiv",
    primaryClass = "astro-ph.HE",
    reportNumber = "INR-TH-2014-015",
    doi = "10.1134/S1063776115040056",
    journal = "J. Exp. Theor. Phys.",
    volume = "120",
    number = "5",
    pages = "790--797",
    year = "2015"
}

@article{Fermi-LAT:2014ryh,
    author = "Ackermann, M. and others",
    collaboration = "Fermi-LAT",
    title = "{The spectrum of isotropic diffuse gamma-ray emission between 100 MeV and 820 GeV}",
    eprint = "1410.3696",
    archivePrefix = "arXiv",
    primaryClass = "astro-ph.HE",
    doi = "10.1088/0004-637X/799/1/86",
    journal = "Astrophys. J.",
    volume = "799",
    pages = "86",
    year = "2015"
}

@article{Murase:2019vdl,
    author = "Murase, Kohta and Kimura, Shigeo S. and Meszaros, Peter",
    title = "{Hidden Cores of Active Galactic Nuclei as the Origin of Medium-Energy Neutrinos: Critical Tests with the MeV Gamma-Ray Connection}",
    eprint = "1904.04226",
    archivePrefix = "arXiv",
    primaryClass = "astro-ph.HE",
    doi = "10.1103/PhysRevLett.125.011101",
    journal = "Phys. Rev. Lett.",
    volume = "125",
    number = "1",
    pages = "011101",
    year = "2020"
}

@article{Inoue:2019fil,
    author = "Inoue, Yoshiyuki and Khangulyan, Dmitry and Inoue, Susumu and Doi, Akihiro",
    title = "{On high-energy particles in accretion disk coronae of supermassive black holes: implications for MeV gamma rays and high-energy neutrinos from AGN cores}",
    journal = "Astrophys.J.",
    volume = "880",
    pages = "40",
    eprint = "1904.00554",
    archivePrefix = "arXiv",
    primaryClass = "astro-ph.HE",
    reportNumber = "RIKEN-iTHEMS-Report-19",
    doi = "10.3847/1538-4357/ab2715",
    month = "4",
    year = "2019"
}

@article{Shakura:1972te,
    author = "Shakura, N. I. and Sunyaev, R. A.",
    title = "{Black holes in binary systems. Observational appearance}",
    journal = "Astron. Astrophys.",
    volume = "24",
    pages = "337--355",
    year = "1973"
}

@inproceedings{Naab:2023xcz,
    author = "Naab, Richard and Ganster, Erik and Zhang, Zelong",
    collaboration = "IceCube",
    title = "{Measurement of the astrophysical diffuse neutrino flux in a combined fit of IceCube's high energy neutrino data}",
    booktitle = "{38th International Cosmic Ray Conference}",
    eprint = "2308.00191",
    archivePrefix = "arXiv",
    primaryClass = "astro-ph.HE",
    reportNumber = "PoS-ICRC2023-1064",
    month = "7",
    year = "2023"
}

@article{Palladino:2020jol,
    author = "Palladino, Andrea and Spurio, Maurizio and Vissani, Francesco",
    title = "{Neutrino Telescopes and High-Energy Cosmic Neutrinos}",
    eprint = "2009.01919",
    archivePrefix = "arXiv",
    primaryClass = "astro-ph.HE",
    doi = "10.3390/universe6020030",
    journal = "Universe",
    volume = "6",
    number = "2",
    pages = "30",
    year = "2020"
}

@article{Loeb:2006tw,
    author = "Loeb, Abraham and Waxman, Eli",
    title = "{The Cumulative background of high energy neutrinos from starburst galaxies}",
    eprint = "astro-ph/0601695",
    archivePrefix = "arXiv",
    doi = "10.1088/1475-7516/2006/05/003",
    journal = "JCAP",
    volume = "05",
    pages = "003",
    year = "2006"
}

@article{Waxman:1997ti,
    author = "Waxman, Eli and Bahcall, John N.",
    title = "{High-energy neutrinos from cosmological gamma-ray burst fireballs}",
    eprint = "astro-ph/9701231",
    archivePrefix = "arXiv",
    reportNumber = "IASSNS-AST-97-14",
    doi = "10.1103/PhysRevLett.78.2292",
    journal = "Phys. Rev. Lett.",
    volume = "78",
    pages = "2292--2295",
    year = "1997"
}

@article{Ptitsyna:2008zs,
    author = "Ptitsyna, Kseniya V. and Troitsky, Sergei V.",
    title = "{Physical conditions in potential sources of ultra-high-energy cosmic rays. I. Updated Hillas plot and radiation-loss constraints}",
    eprint = "0808.0367",
    archivePrefix = "arXiv",
    primaryClass = "astro-ph",
    doi = "10.3367/UFNe.0180.201007c.0723",
    journal = "Phys. Usp.",
    volume = "53",
    pages = "691--701",
    year = "2010"
}

@article{Hillas:1984ijl,
    author = "Hillas, A. M.",
    title = "{The Origin of Ultrahigh-Energy Cosmic Rays}",
    doi = "10.1146/annurev.aa.22.090184.002233",
    journal = "Ann. Rev. Astron. Astrophys.",
    volume = "22",
    pages = "425--444",
    year = "1984"
}

@article{Sotirov:2022uqk,
    author = "Sotirov, Simon",
    title = "{General constraints on sources of high-energy cosmic rays from interaction losses}",
    eprint = "2212.03483",
    archivePrefix = "arXiv",
    primaryClass = "astro-ph.HE",
    reportNumber = "INR-TH-2022-027",
    doi = "10.1103/PhysRevD.107.123018",
    journal = "Phys. Rev. D",
    volume = "107",
    number = "12",
    pages = "123018",
    year = "2023"
}

@article{Fabian:1992rt,
    author = "Fabian, A. C. and Barcons, X.",
    title = "{The origin of the x-ray background}",
    doi = "10.1146/annurev.aa.30.090192.002241",
    journal = "Ann. Rev. Astron. Astrophys.",
    volume = "30",
    pages = "429--456",
    year = "1992"
}

@article{Ueda:2014tma,
    author = {Ueda, Yoshihiro and Akiyama, Masayuki and Hasinger, G\"unther and Miyaji, Takamitsu and Watson, Michael G.},
    title = "{Toward the Standard Population Synthesis Model of the X-Ray Background: Evolution of X-Ray Luminosity and Absorption Functions of Active Galactic Nuclei Including Compton-Thick Populations}",
    eprint = "1402.1836",
    archivePrefix = "arXiv",
    primaryClass = "astro-ph.CO",
    doi = "10.1088/0004-637X/786/2/104",
    journal = "Astrophys. J.",
    volume = "786",
    pages = "104",
    year = "2014"
}

@article{Ajello:2008xb,
    author = "Ajello, M. and others",
    title = "{Cosmic X-ray background and Earth albedo Spectra with Swift/BAT}",
    eprint = "0808.3377",
    archivePrefix = "arXiv",
    primaryClass = "astro-ph",
    doi = "10.1086/592595",
    journal = "Astrophys. J.",
    volume = "689",
    pages = "666",
    year = "2008"
}

@article{Hasinger:2005sb,
    author = "Hasinger, Gunther and Miyaji, Takamitsu and Schmidt, Maarten",
    title = "{Luminosity-dependent evolution of soft x-ray selected AGN: New Chandra and XMM-Newton surveys}",
    eprint = "astro-ph/0506118",
    archivePrefix = "arXiv",
    doi = "10.1051/0004-6361:20042134",
    journal = "Astron. Astrophys.",
    volume = "441",
    pages = "417--434",
    year = "2005"
}

@inproceedings{Karas:2019lju,
    author = "Karas, Vladimir and Svoboda, Jiri and Zajacek, Michal",
    title = "{Selected Chapters on Active Galactic Nuclei as Relativistic Systems}",
    eprint = "1901.06507",
    archivePrefix = "arXiv",
    primaryClass = "astro-ph.HE",
    month = "1",
    year = "2019"
}

@article{Kompaneets:1957,
    author = "A.S. Kompaneets",
    title = "{The Establishment of Thermal Equilibrium between Quanta and Electrons}",
    journal = "” Sov. Phys. JETP",
    volume = "4",
    pages = "730",
    year = "1957"
}

@article{Padovani:2024ibi,
    author = "Padovani, P. and others",
    title = "{High-energy neutrinos from the vicinity of the supermassive black hole in NGC\,1068}",
    eprint = "2405.20146",
    archivePrefix = "arXiv",
    primaryClass = "astro-ph.HE",
    doi = "10.1038/s41550-024-02339-z",
    journal = "Nature Astron.",
    volume = "8",
    number = "9",
    pages = "1077--1087",
    year = "2024"
}

@article{Gianolli:2023zji,
    author = "Gianolli, V. E. and others",
    title = "{Uncovering the geometry of the hot X-ray corona in the Seyfert galaxy NGC 4151 with IXPE}",
    eprint = "2303.12541",
    archivePrefix = "arXiv",
    primaryClass = "astro-ph.GA",
    doi = "10.1093/mnras/stad1697",
    journal = "Mon. Not. Roy. Astron. Soc.",
    volume = "523",
    number = "3",
    pages = "4468--4476",
    year = "2023"
}

@article{Tagliacozzo:2023apk,
    author = "Tagliacozzo, D. and others",
    title = "{The geometry of the hot corona in MCG-05-23-16 constrained by X-ray polarimetry}",
    eprint = "2305.10213",
    archivePrefix = "arXiv",
    primaryClass = "astro-ph.HE",
    doi = "10.1093/mnras/stad2627",
    journal = "Mon. Not. Roy. Astron. Soc.",
    volume = "525",
    number = "3",
    pages = "4735--4743",
    year = "2023"
}

@article{Ingram:2023twz,
    author = "Ingram, A. and others",
    title = "{The X-ray polarization of the Seyfert 1 galaxy IC 4329A}",
    eprint = "2305.13028",
    archivePrefix = "arXiv",
    primaryClass = "astro-ph.HE",
    doi = "10.1093/mnras/stad2625",
    journal = "Mon. Not. Roy. Astron. Soc.",
    volume = "525",
    number = "4",
    pages = "5437--5449",
    year = "2023"
}

@article{Abbasi:2024ofy,
    author = "Abbasi, R. and others",
    title = "{IceCube Search for Neutrino Emission from X-ray Bright Seyfert Galaxies}",
    journal = "",
    eprint = "2406.07601",
    archivePrefix = "arXiv",
    primaryClass = "astro-ph.HE",
    month = "6",
    year = "2024"
}

@ARTICLE{Pozd:1979ast,
    author = {{Pozdniakov}, L.~A. and {Sobol}, I.~M. and {Siuniaev}, R.~A.},
    title = "{Comptonization and radiation spectra of X-ray sources. Monte Carlo computations.}",
    journal = {Pisma v Astronomicheskii Zhurnal},
    year = 1979,
    volume = {5},
    pages = {279-284},
}

@ARTICLE{Pozd:1983sun,
    author = {{Pozdnyakov}, L.~A. and {Sobol}, I.~M. and {Syunyaev}, R.~A.},
    title = "{Comptonization and the shaping of X-ray source spectra - Monte Carlo calculations}",
    journal = {Sov. Sci. Rev., Section E: Astro. and Space Phys. Rev.},
    year = 1983,
    volume = {2},
    pages = {189-331},
}

@article{Ueda:2003yx,
    author = "Ueda, Yoshihiro and Akiyama, Masayuki and Ohta, Kouji and Miyaji, Takamitsu",
    title = "{Cosmological evolution of the hard x-ray AGN luminosity function and the origin of the hard x-ray background}",
    eprint = "astro-ph/0308140",
    archivePrefix = "arXiv",
    doi = "10.1086/378940",
    journal = "Astrophys. J.",
    volume = "598",
    pages = "886--908",
    year = "2003"
}

@article{Levinson,
  title = {Particle Acceleration and Curvature TeV Emission by Rotating, Supermassive Black Holes},
  author = {Levinson, Amir},
  journal = {Phys. Rev. Lett.},
  volume = {85},
  issue = {5},
  pages = {912--915},
  numpages = {0},
  year = {2000},
  doi = {10.1103/PhysRevLett.85.912},
}

@article{Neronov:2009zz,
    author = "Neronov, A. Yu and Semikoz, D. V. and Tkachev, I. I.",
    title = "{Ultra-High Energy Cosmic Ray production in the polar cap regions of black hole magnetospheres}",
    eprint = "0712.1737",
    archivePrefix = "arXiv",
    primaryClass = "astro-ph",
    doi = "10.1088/1367-2630/11/6/065015",
    journal = "New J. Phys.",
    volume = "11",
    pages = "065015",
    year = "2009"
}

@article{Ptitsyna:2015nta,
    author = "Ptitsyna, K. and Neronov, A.",
    title = "{Particle acceleration in the vacuum gaps in black hole magnetospheres}",
    eprint = "1510.04023",
    archivePrefix = "arXiv",
    primaryClass = "astro-ph.HE",
    doi = "10.1051/0004-6361/201527549",
    journal = "Astron. Astrophys.",
    volume = "593",
    pages = "A8",
    year = "2016"
}

@article{IceCube:2022der,
    author = "Abbasi, R. and others",
    collaboration = "IceCube",
    title = "{Evidence for neutrino emission from the nearby active galaxy NGC 1068}",
    eprint = "2211.09972",
    archivePrefix = "arXiv",
    primaryClass = "astro-ph.HE",
    doi = "10.1126/science.abg3395",
    journal = "Science",
    volume = "378",
    number = "6619",
    pages = "538--543",
    year = "2022"
}

@article{IceCube:2018cha,
    author = "Aartsen, M. G. and others",
    collaboration = "IceCube",
    title = "{Neutrino emission from the direction of the blazar TXS 0506+056 prior to the IceCube-170922A alert}",
    eprint = "1807.08794",
    archivePrefix = "arXiv",
    primaryClass = "astro-ph.HE",
    doi = "10.1126/science.aat2890",
    journal = "Science",
    volume = "361",
    number = "6398",
    pages = "147--151",
    year = "2018"
}

@article{Plavin:2020mkf,
    author = "Plavin, A. V. and Kovalev, Y. Y. and Kovalev, Yu. A. and Troitsky, S. V.",
    title = "{Directional Association of TeV to PeV Astrophysical Neutrinos with Radio Blazars}",
    eprint = "2009.08914",
    archivePrefix = "arXiv",
    primaryClass = "astro-ph.HE",
    reportNumber = "INR-TH-2020-039",
    doi = "10.3847/1538-4357/abceb8",
    journal = "Astrophys. J.",
    volume = "908",
    number = "2",
    pages = "157",
    year = "2021"
}

@article{McDonough:2023ngk,
    author = "McDonough, K. and Hughes, K. and Smith, D. and Vieregg, A. G.",
    title = "{A search for AGN sources of the IceCube diffuse neutrino flux}",
    eprint = "2307.04194",
    archivePrefix = "arXiv",
    primaryClass = "astro-ph.HE",
    doi = "10.1088/1475-7516/2024/06/035",
    journal = "JCAP",
    volume = "06",
    pages = "035",
    year = "2024"
}

@article{Kovalev:2022izi,
    author = "Kovalev, Y. Y. and Plavin, A. V. and Troitsky, S. V.",
    title = "{Galactic Contribution to the High-energy Neutrino Flux Found in Track-like IceCube Events}",
    eprint = "2208.08423",
    archivePrefix = "arXiv",
    primaryClass = "astro-ph.HE",
    reportNumber = "INR-TH-2022-018",
    doi = "10.3847/2041-8213/aca1ae",
    journal = "Astrophys. J. Lett.",
    volume = "940",
    number = "2",
    pages = "L41",
    year = "2022"
}

@article{IceCube:2023ame,
    author = "Abbasi, R. and others",
    collaboration = "IceCube",
    title = "{Observation of high-energy neutrinos from the Galactic plane}",
    eprint = "2307.04427",
    archivePrefix = "arXiv",
    primaryClass = "astro-ph.HE",
    doi = "10.1126/science.adc9818",
    journal = "Science",
    volume = "380",
    number = "6652",
    pages = "adc9818",
    year = "2023"
}

@article{Baikal-GVD:2024kfx,
    author = "Allakhverdyan, V. A. and others",
    collaboration = "Baikal-GVD",
    title = "{Probing the Galactic Neutrino Flux at Neutrino Energies above 200 TeV with the Baikal Gigaton Volume Detector}",
    eprint = "2411.05608",
    archivePrefix = "arXiv",
    primaryClass = "astro-ph.HE",
    doi = "10.3847/1538-4357/adb630",
    journal = "Astrophys. J.",
    volume = "982",
    number = "2",
    pages = "73",
    year = "2025"
}

@article{Neronov:2025cfc,
    author = "Neronov, A. and Kalashev, O. and Semikoz, D. V. and Savchenko, D. and Poleshchuk, M.",
    title = "{Neutrino emission and corona heating induced by high-energy proton interactions in Seyfert galaxies}",
    eprint = "2503.16273",
    archivePrefix = "arXiv",
    primaryClass = "astro-ph.HE",
    year = "2025",
    journal=""
}

@article{Blandford:1987pw,
    author = "Blandford, R. and Eichler, D.",
    title = "{Particle Acceleration at Astrophysical Shocks: A Theory of Cosmic Ray Origin}",
    doi = "10.1016/0370-1573(87)90134-7",
    journal = "Phys. Rept.",
    volume = "154",
    pages = "1--75",
    year = "1987"
}

@article{Zhdankin:2018lhq,
    author = "Zhdankin, Vladimir and Uzdensky, Dmitri A. and Werner, Gregory R. and Begelman, Mitchell C.",
    title = "{Electron and ion energization in relativistic plasma turbulence}",
    eprint = "1809.01966",
    archivePrefix = "arXiv",
    primaryClass = "astro-ph.HE",
    doi = "10.1103/PhysRevLett.122.055101",
    journal = "Phys. Rev. Lett.",
    volume = "122",
    number = "5",
    pages = "055101",
    year = "2019"
}

@article{Hoshino:2013pza,
    author = "Hoshino, Masahiro",
    title = "{Particle Acceleration during Magnetorotational Instability in a Collisionless Accretion Disk}",
    eprint = "1306.6720",
    archivePrefix = "arXiv",
    primaryClass = "astro-ph.HE",
    doi = "10.1088/0004-637X/773/2/118",
    journal = "Astrophys. J.",
    volume = "773",
    pages = "118",
    year = "2013"
}

@article{Murase:2015xka,
    author = "Murase, Kohta and Guetta, Dafne and Ahlers, Markus",
    title = "{Hidden Cosmic-Ray Accelerators as an Origin of TeV-PeV Cosmic Neutrinos}",
    eprint = "1509.00805",
    archivePrefix = "arXiv",
    primaryClass = "astro-ph.HE",
    doi = "10.1103/PhysRevLett.116.071101",
    journal = "Phys. Rev. Lett.",
    volume = "116",
    number = "7",
    pages = "071101",
    year = "2016"
}

@article{Kisaka:2020lfl,
    author = "Kisaka, Shota and Levinson, Amir and Toma, Kenji",
    title = "{Comprehensive analysis of magnetospheric gaps around Kerr black holes using 1D GRPIC simulations}",
    eprint = "2007.02838",
    archivePrefix = "arXiv",
    primaryClass = "astro-ph.HE",
    doi = "10.3847/1538-4357/abb46c",
    journal = "Astrophys. J.",
    volume = "902",
    number = "1",
    pages = "80",
    year = "2020"
}

@article{Zdziarski:2019cvs,
    author = "Zdziarski, Andrzej A. and Szanecki, Michal and Poutanen, Juri and Gierlinski, Marek and Biernacki, Pawel",
    title = "{Spectral and temporal properties of Compton scattering by mildly relativistic thermal electrons}",
    eprint = "1910.04535",
    archivePrefix = "arXiv",
    primaryClass = "astro-ph.HE",
    doi = "10.1093/mnras/staa159",
    journal = "Mon. Not. Roy. Astron. Soc.",
    volume = "492",
    number = "4",
    pages = "5234--5246",
    year = "2020"
}

@article{Lee:1996fp,
    author = "Lee, Sangjin",
    title = "{On the propagation of extragalactic high-energy cosmic and gamma-rays}",
    eprint = "astro-ph/9604098",
    archivePrefix = "arXiv",
    reportNumber = "FERMILAB-PUB-96-066-A",
    doi = "10.1103/PhysRevD.58.043004",
    journal = "Phys. Rev. D",
    volume = "58",
    pages = "043004",
    year = "1998"
}

@article{Finke:2008pe,
    author = {Finke, Justin D. and Dermer, Charles D. and B{\"o}ttcher, Markus},
    title = "{Synchrotron Self-Compton Analysis of TeV X-ray Selected BL Lacertae Objects}",
    eprint = "0802.1529",
    archivePrefix = "arXiv",
    primaryClass = "astro-ph",
    doi = "10.1086/590900",
    journal = "Astrophys. J.",
    volume = "686",
    pages = "181",
    year = "2008"
}

@article{Inoue:2016fwn,
    author = "Inoue, Yoshiyuki and Tanaka, Yasuyuki T.",
    title = "{Baryon Loading Efficiency and Particle Acceleration Efficiency of Relativistic Jets: Cases For Low Luminosity BL Lacs}",
    eprint = "1603.07623",
    archivePrefix = "arXiv",
    primaryClass = "astro-ph.HE",
    doi = "10.3847/0004-637X/828/1/13",
    journal = "Astrophys. J.",
    volume = "828",
    number = "1",
    pages = "13",
    year = "2016"
}

@article{Kimura:2018clk,
    author = "Kimura, Shigeo S. and Tomida, Kengo and Murase, Kohta",
    title = "{Acceleration and Escape Processes of High-energy Particles in Turbulence inside Hot Accretion Flows}",
    eprint = "1812.03901",
    archivePrefix = "arXiv",
    primaryClass = "astro-ph.HE",
    doi = "10.1093/mnras/stz329",
    journal = "Mon. Not. Roy. Astron. Soc.",
    volume = "485",
    number = "1",
    pages = "163--178",
    year = "2019"
}


\end{document}